\documentclass[runningheads]{llncs}
\pdfoutput=1

\usepackage[T1]{fontenc}
\usepackage{graphicx}

\usepackage[colorlinks=true,allcolors=black]{hyperref}
\usepackage{amsmath}
\usepackage{cleveref}
\usepackage{lipsum}
\usepackage{listings}
\usepackage{color}

\usepackage{hyperref}
\usepackage{bookmark}

\usepackage{todonotes}
\newcommand{\comment}[1]{}

\begin{document}

\title{Finding Software Supply Chain Attack Paths with Logical Attack Graphs}
\author{Luís Soeiro\inst{1}\orcidID{0009-0003-8609-1352} \and
	Thomas Robert\inst{1}\orcidID{0000-0002-4423-5720} \and
	Stefano Zacchiroli\inst{1}\orcidID{0000-0002-4576-136X}}
\authorrunning{L. Soeiro et al.}

\institute{LTCI, Télécom Paris, Institut Polytechnique
	de Paris, France
	\url{https://www.ip-paris.fr} \\
	\email{\{luis.soeiro,thomas.robert,stefano.zacchiroli\}@telecom-paris.fr}}

\maketitle              \begin{abstract}
Cyberattacks are becoming increasingly frequent and sophisticated,
	often exploiting the software supply chain (SSC) as an attack vector.
	Attack graphs provide a detailed representation of the sequence of
	events and vulnerabilities that could lead to a successful security
	breach in a system. MulVal is a widely used open-source tool for
	logical attack graph generation in networked systems. However, its
	current lack of support for capturing and reasoning about SSC threat
	propagation makes it unsuitable for addressing modern SSC attacks, such
	as the XZ compromise or the 3CX double SSC attack. To address this
	limitation, we propose an extension to MulVal that integrates SSC
	threat propagation analysis with existing network-based threat
	analysis. This extension introduces a new set of predicates within the
	familiar MulVal syntax, enabling seamless integration. The new facts
	and interaction rules model SSC assets, their dependencies,
	interactions, compromises, additional security mechanisms, initial
	system states, and known threats. We explain how this integration
	operates in both directions and demonstrate the practical application
	of the extension. 
\end{abstract}

\section{Introduction}

Advances in information technology have consistently been shadowed by
the proliferation and rising complexity of cyberattacks. The widespread
adoption of Free and Open Source Software (FOSS), driven by scientific,
industrial, and economic motivations~\cite{Paschali2017}, has further
expanded the attack surface due to its distributed and
resource-constrained development model~\cite{Ladisa2023a}. Within this
context, the Software Supply Chain (SSC) has emerged as a critical
target. Its global interconnectedness and limited transparency enable
threat actors to exploit vulnerabilities (e.g., Log4Shell) or introduce
malicious code, bypassing traditional defenses and propagating attacks
across dependent systems~\cite{Williams2025}.

SSC attacks can also be combined. In the 2023 case of the 3CX
attack~\cite{r_3cx}, two SSC attacks had to be carried out in a
sequence of events. First, the server that distributed Trading
Technologies' software was compromised, leading to the injection of a
backdoor in the X\_Trader software, which was then available for
download. Then, an employee of the 3CX company downloaded the X\_Trader
software and executed it on his personal computer. The malicious
software then helped threat actors connect to the 3CX systems using the
employee's authenticated VPN connection. The attackers ultimately
compromised the 3CX build environment, injecting malicious code into
the signed Windows and macOS versions of the 3CXDesktopApp, which
affected the company’s customers.

Many models exist to capture threat knowledge and the progression of
attacks on a network. Attack trees and attack graphs are used to
decompose and understand the steps involved in complex attack
scenarios~\cite{Lallie2020}. Moreover, such models can be automatically
generated from system introspection or from Cyber Threat Intelligence
streams of data~\cite{Konsta2024}. While attack trees capture a single
attack goal, attack graphs can capture multiple attack
goals~\cite{SaintHilaire2024} and multiple attack
paths~\cite{Konsta2024}. The Logical Attack Graph (LAG) formalism
introduced in the seminal work of MulVal~\cite{Ou2005} is widely used
and has been regularly extended over the past two
decades~\cite{Tayouri2023}. However, neither MulVal nor its extensions
are prepared to reason about SSC threats~\cite{Duman2024}. This paper
introduces a new MulVal extension that integrates SSC threat
propagation reasoning. A full replication package containing all the
code presented in this work is available from Zenodo~\cite{r_zenodo}.

\textit{The following research questions will be answered in this work:}
\label{RQ}

\textbf{RQ1:} \textit{To what extent is it possible to formalize knowledge of SSC attacks into LAG?}

\textbf{RQ2:} \textit{To what extent does such a formalism uncover non-trivial
	attack scenarios?}

This paper is organized as follows: \Cref{sec:rw} presents related
work; \Cref{sec:background} provides background on MulVal;
\Cref{sec:extension} introduces our contribution and approach;
\Cref{sec:threat-propagation} details how the extension rules integrate
with MulVal; \Cref{sec:usage} presents scenarios demonstrating
real-world use; \Cref{sec:discussion} revisits the research questions;
\Cref{sec:limitations} discusses limitations and possible mitigations;
and \Cref{sec:conclusion} concludes with closing remarks and future
work.

\section{Related work}
\label{sec:rw}

There is substantial research on SSC attack and countermeasure
elicitation~\cite{Ladisa2023a}, including work on malware enabling such
attacks~\cite{Ohm2020,Martinez2021} and SSC technical
processes~\cite{Hammi2023}. However, these analyses cover only the SSC
assets without clear links to the systems that depend on them. The log
model~\cite{Soeiro2023} proposes threat propagation reasoning for the
SSC, but it lacks modeling of available security mechanisms, making its
analysis pessimistic, and it does not address the extra complexity of
modeling cyberattacks against networked systems. Attack trees and
attack graphs generalize complex scenarios~\cite{Lallie2020} and have
been applied to SSCs~\cite{Ladisa2023}; attack graphs (LAGs) better
support multiple goals and paths~\cite{Konsta2024}. To our knowledge
only two works attempt to bridge SSC and networked‑system scopes: the
Hardening Framework for Substations offers interactive countermeasures
but models Software Supply Chain Attacks (SSCA) as a simple Boolean
state~\cite{Duman2024}, and CORAL extends MulVal LAGs for container
risks~\cite{Tayouri2025} yet does not capture the full range of SSCA
tampering scenarios.

MulVal cannot compute SSC threat propagation because it lacks
predicates for SSC graphs and propagation rules, requires a priori
vulnerability declarations (so emergent paths are missed), cannot model
vulnerabilities that enable unintended network connections, and has no
malicious-software constructs.

\section{Background}
\label{sec:background}

MulVal is a widely used open-source tool for generating logical attack
graphs for networked systems. In these graphs, nodes represent logical
statements about system state or attacker capabilities. MulVal models
and reasons about those statements using Datalog, a declarative logic
language (a safe subset of Prolog) for defining and querying deductive
databases. It relies on five concepts: variables, constants,
predicates, formulae, facts and inference rules. A predicate formula is
an expression $pred_1(t_1,\dots,t_n)$ where $pred_1$ is the predicate
name and $t_i$ are the terms it applies to. A formula declared true is
a \textsl{fact}; otherwise it is used to define \textsl{inference
	rules}. An inference rule is a Horn clause, $P_0:-P_1,\dots,P_n$, such
that the predicate formula $P_0$ is true when the conjunction of $P_1
	\wedge \dots \wedge P_n$ is true. The terms used in the predicate
formula can be constants (strings used as identifiers of the problem
modeled) or variables. Variables are used to describe the constraints
binding the parameters among $P_0,\dots,P_n$. The deductive database is
the combination of the facts and all the inference rules. It can be
queried to determine if some predicate formulae are true. The
interpreter of these queries can provide the trace of all the rules
used. MulVal encodes system state as facts and attacker behaviors as
inference rules, so analyses produce attack-graph derivations that show
exactly which facts and rules lead to a compromise.

\section{A MulVal extension for SSC}
\label{sec:extension}

We extend MulVal to capture and reason about the core elements of the
SSC graph: \label{elements} \textit{host (H)}, \textit{build
	environment (BE)}, \textit{transformer (T)}, and \textit{software
	artifact (SA)}~\cite{Soeiro2023}. These assets of the SSC depend on
each other. The dependencies define an SSC graph, in which the assets
are the vertices.

\subsection{Approach}

In this section, we introduce new predicates to capture SSC assets and
their dependencies. Then, we introduce new predicates to capture some
previously uncovered aspects of attack behavior. Finally, we introduce
predicates and inference rules to capture security mechanisms in the
SSC. This work paves the way for \Cref{sec:threat-propagation}, which
presents their integration into MulVal and the resulting threat
propagation analysis framework.

To compute the SSC contributions to the threat level on a given vertex
$e$ of the SSC graph we need to trace the contributions of all other
vertices on paths that reach $e$. For instance, the threat level of a
\textit{software artifact} $sa$ depends on the threat levels of all the
SSC vertices that are on paths leading to $sa$, e.g., hosts, build
environments, and other software artifacts used as input. Conversely,
taking advantage of tools like MulVal to capture usual attacks (e.g.,
principal compromise, vulnerability exploit) on hosts (either virtual
or physical) contributes to a better assessment of threats for those
vertices in the SSC.

\subsection{Modeling SSC assets and their interactions}
\label{sec:ssc-assets}

The SSC assets (i.e., the SSC-graph core elements), their dependency
organization, and their initial known unsafe state (i.e., vulnerable or
compromised) are modeled by the newly introduced predicates:

\renewcommand{\topfraction}{.8}
\renewcommand{\floatpagefraction}{.8}

\begin{figure}
	\centering
\includegraphics[width=0.9\textwidth]{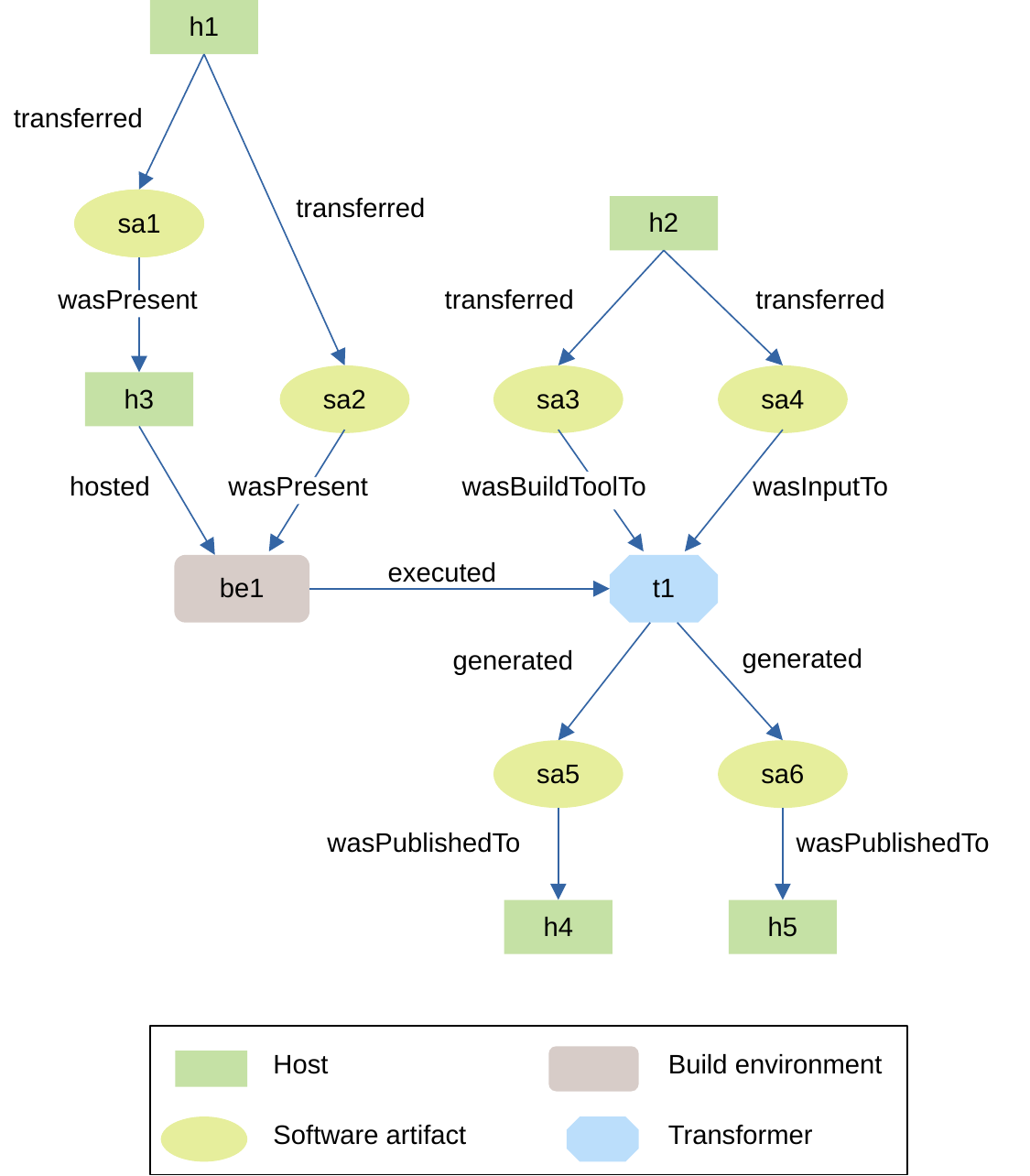}
	\caption{The software supply chain for software artifacts sa5 e sa6}
	\label{fig:ssc-1}
\end{figure}

\begin{itemize}
	\item \texttt{vulNetworkProperty(vulID, protocol, port, user)} -- binds a
	      vulnerability identifier \textit{vulID} to a network protocol, port,
	      and user. This enables the system to model the case where a
	      vulnerability transforms a piece of software not intended to
	      provide a network service into an access point for a remote attacker
	      (e.g., exposure of the \textit{RMI} protocol on port 1099 in the
	      Log4Shell vulnerability~\cite{Everson2022});
	\item \texttt{signed<X>($key$, $e$)}, $X \in \{C, SA\}$ -- declares that
	      $key$ was used to sign certificate or software artifact $e$.
	\item \texttt{issued($Cert_1$, $Cert_2$)} -- declares that
	      $Cert_1$ has been used to issue $Cert_2$;
	\item \texttt{compromised<X>($e$)}, $X \in \{H, BE, T, K, C\}$ --
	      declares that a given element of type $X$ is compromised. $H$, $BE$,
	      $T$, $K$, and $C$ denote, respectively, \textit{host},
	      \textit{build environment}, \textit{transformer},
	      \textit{signing key}, and \textit{certificate};
	\item \texttt{maliciousSA($sa$)} -- declares that \textit{software artifact}
	      $sa$ is known to be malicious;
	\item \texttt{isolationEscapeBE($BE$)} -- used to infer situations where
	      there is an escape from an isolation mechanism;
	\item \texttt{hosted($h$,$be$)}, \texttt{executed($be$,$t$)},
	      \texttt{wasInputTo($sa$,$t$)}, \texttt{wasBuildToolTo($sa$,$t$)},
	      \texttt{wasPresent($sa$,$h$)}, \texttt{generated($t$,$sa$)},
	      \texttt{wasPublishedTo($sa$,$h$)}, and \texttt{trans\-ferred($sa$, $h$)},
	      with $sa$, $h$, $be$, and $t$ denoting software artifact, host,
	      build environment (where software builds occur), and transformer
	      (the set of operations that take software artifacts and build tools
	      as input and generate new software artifacts), respectively --
	      derived from the Log Model edges~\cite{Soeiro2023}.
	      \Cref{fig:ssc-1} shows an example of an SSC graph that contains all edge types and all elements.
\end{itemize}

\subsection{Malicious software artifacts}
\label{sec:malicous-software}

It has been observed that the complexity (e.g., lines of code, number
of source files) of malicious software increases roughly at one order
of magnitude per decade~\cite{Calleja2016}. The number of malicious
software artifacts being uploaded to popular programming-language
repositories (e.g., PyPI, CRAN, npm) has surpassed the number of
vulnerable software~\cite{Ruohonen2025}. While malicious software can
have many different behaviors~\cite{Lindorfer2012}, they share one
common trait: the ability to autonomously perform actions. That changes
the way attack progress is usually modeled in MulVal. Thus, we
introduce the \texttt{execBatchCode} predicate that models this attack
behavior to MulVal's reasoning system. Similarly to \texttt{execCode},
we define interaction rules that state the conditions under which the
system will autonomously execute malicious code. Listing
\ref{lst:exec-batch} shows one \texttt{execBatchCode} interaction rule.
If a software artifact \textit{SA} was observed executing on host
\textit{Host}, was classified as malicious (determined via
\texttt{maliciousSA(SA)}), and executed under principal $User$ (with
\texttt{canAccessFile(...)} used to determine the principal under which
$SA$ ran), then $SA$ can autonomously execute code as principal
\textit{User} on host \textit{Host} without human interaction.

\begin{lstlisting}[float,basicstyle=\small,caption={Interaction rule that shows one effect of a malicious software artifact},label=lst:exec-batch]
  execBatchCode(Host, SA, User) :-
    wasPresent(SA, Host), 
    maliciousSA(SA),
    canAccessFile(Host, User, Access, SA) 
\end{lstlisting}

\subsection{Modeling security mechanisms in SSC}

Security mechanisms act as countermeasures for attacks (e.g., a working
firewall prevents an outside connection to a vulnerable internal
service). In the absence of a reachable privilege-escalation
vulnerability or credential theft, the operating-system access-control
mechanism prevents a malicious software artifact running as one
principal from injecting code into another artifact running as a
different principal. When computing possible attack paths for a
scenario we consider the preventive nature of the existing security
mechanisms that are deployed. This consideration makes the
threat-propagation analysis more accurate by removing unreachable
attack paths.

MulVal already models many security mechanisms, and our extension takes
advantage of them. However, two mechanisms broadly used in SSC are
missing in MulVal: build-environment isolation and authenticity
verification of software artifacts.

\paragraph{Isolation of build environments}

We consider existing security mechanisms that can prevent the
propagation of threats. Isolation of a build environment prevents the
flow of threats from it or to it. For instance, let a \textit{host}
$H_1$ provide isolation for its \textit{build environments} $BE_1$ and
$BE_2$. Then a malicious \textit{software artifact} running in $BE_1$
cannot propagate malicious code to the $BE_2$ asset or to assets that
depend on it. Yet, if the isolation mechanism is compromised or if
dependencies exist between the assets in $BE_2$ and those produced in
$BE_1$, propagation can still occur. We define predicates to cover a
reasonable set of cases.

\begin{lstlisting}[float,basicstyle=\small,caption={Interaction rule that shows the conditions for an escape of the build-environment isolation},label=lst:be-escape]
isolationEscapeBE(BE) :- 
  execBatchCode(BE, SA, User),
  wasPresent(VulnSA, BE),
  vulExists(BE, _, VulnSA, localExploit, isolationEscape)

isolationEscapeBE(BE) :- 
  execBatchCode(BE, SA, User),
  hosted(H, BE),
  wasPresent(VulnSA, H),
  vulExists(H, _, VulnSA, localExploit, isolationEscape)
\end{lstlisting}

Different isolation mechanisms for computer systems are available
(e.g., processes, containers, virtualization), each with trade-offs
between security and performance overheads~\cite{Shu2016}.
Independently of the underling isolation mechanism, we model the
possible isolation outcomes using the concepts of isolated build
environment and MulVal access control. We show two scenarios. In the
first, there is no isolation of the \textit{build environment} $be_1$;
only access control is used to prevent one build run from interfering
with other build runs on the same \textit{host} $h_1$. In this case we
model it by declaring a single $be_1$ in $h_1$, \texttt{hosted($h_1$,
	$be_1$)}, and one \textit{transformer} $t_i$ for each build run that is
executed. Let $N$ be the number of independent builds. We declare the
predicates \texttt{executed($be_1$, $t_i$)} and
\texttt{localFileProtection($be_1$, $user_i$, $access_i$, ${pSA}_i$)}
for $i \in \{1 \dots N\}$, where $user_i$, $access_i$, and ${pSA}_i$
are, respectively, the principal, the access type (e.g., read, write),
and the logical path of the \textit{software artifacts} used by each
build run.

The second scenario aligns with current expectations of isolation that
come from using build platforms. Users are relying more on services
that offer Continuous Integration/Continuous Deployment (CI/CD)
workflows for building software~\cite{Mazrae2023}. In this scenario,
each build environment is isolated from the others. We model it by
declaring multiple \textit{build environments}, each with only one
\textit{transformer}. Let $h_1$ be the host and $N$ be the number of
independent builds. We declare the predicates \texttt{hosted($h_1$,
	$be_i$)} and \texttt{executed($be_i$, $t_i$)} for $i \in \{1 \dots
	N\}$. We assume that each $be_i$ is isolated.

Since vulnerabilities may allow for process escape (i.e., privilege
escalation), container escape~\cite{Lin2018} or virtualization
escape~\cite{Pearce2013}, we add new rules to MulVal to capture those
interactions in the context of the SSC. We define the predicate
\texttt{isolationEscapeBE($BE$)} to cover both virtualization and
container escape. Listing \ref{lst:be-escape} shows two interaction
rules that allow modeling the situation where a malicious software
artifact escapes from the isolating container or virtual machine. The
first rule is triggered by a vulnerable software artifact located
inside the \textit{build environment} (container or virtual machine)
and the second rule is triggered by an artifact located on the
\textit{host} that hosted the \textit{build environment}. For both
escapes to succeed, there must be a software artifact that has received
the propagation of a vulnerability with the property
\texttt{vulProperty(vulID, localExploit, isolationEscape)}.

\paragraph{Authenticity of software artifacts}

Several SSC attack paths rely on hijacking the secure dissemination of
software~\cite{Ladisa2023}. To improve software dissemination,
distribution systems either started to rely on digital
signatures~\cite{Catuogno2020} or proposed
them~\cite{kalu2024,Duman2024}. Modeling authentication mechanisms
within the SSC is complex; to keep the analysis tractable, we model
attacks limited to software-artifact tampering and trust-chain
compromises. Because signing adoption varies
widely~\cite{Schorlemmer2024}, we account for cases where data
authenticity is enforced or absent for different objects. For example,
a system may obtain software from official repositories, verified by
the operating-system package manager, or from unverified sources (e.g.,
PyPI).

Data authentication relies heavily on certificates and signing keys.
They build a trust chain from a root of trust through root anchors up
to the certificate used to validate a software artifact. Yet, threat
actors are compromising code-signing mechanisms to distribute malicious
software as legitimate (e.g., XZ, SolarWinds, 3CX)~\cite{Ji2024}. We
introduce rules to identify the effects of key compromises at any stage
of trust chains. Modeling certificate chains is done through the
predicate \texttt{issued($Cert_1$, $Cert_2$)}. It captures trust
dependencies along the chain. We identify keys and signed objects with
the predicates \texttt{signedSA($Key$, $Sa$)} (signing \textit{software
	artifacts}) and \texttt{signedC($Key$, $Cert$)} (signing
\textit{certificates}). This allows the rules to identify the effects
of key compromises at any stage of the trust chain. We introduce the
predicate \texttt{validateSA($Cert$, $Sa$)} to declare that
authenticity is checked for $Sa$ using the public key bound to $Cert$
along the SSC (e.g., for all packages from a Debian GNU/Linux
distribution). These predicates are sufficient to cover the basics of
SSC data-authenticity mechanisms.

\begin{lstlisting}[float,basicstyle=\small,caption={Interaction rule that shows SA vulnerability propagation},label=lst:signing]
compromisedK(PrivateKey) :-
  compromisedH(H),
  wasPresent(PrivateKey, H)

compromisedC(Certificate) :-
  compromisedK(PrivateKey),
  signedC(PrivateKey, Certificate)  

maliciousSA(SA) :- 
  compromisedC(Certificate),
  validateSA(Certificate, SA)
\end{lstlisting}

Compromises of private keys or corresponding certificates are defined
using the predicates \texttt{compromisedK($key$)} and
\texttt{compromisedC($Cert$)}. Our extension then considers all
\textit{software artifacts} that were signed by a compromised private
key as malicious and propagates the consequences. Listing
\ref{lst:signing} shows some of the rules that account for violations
of privacy or integrity of signing keys. The first rule states that the
private key $PrivateKey$ is compromised if it was stored on a
compromised \textit{host}. The second rule states that a certificate
signed by a compromised key is also compromised. The third rule states
that a software artifact signed with a compromised key is compromised.

\section{Integration of SSC threat propagation with MulVal}
\label{sec:threat-propagation}

We introduced new predicates to capture SSC assets, their dependencies,
and their threat states. Yet, we need to introduce new predicates that
bridge our new rules and MulVal's existing rules.

\subsection{Vulnerable software propagation}

\begin{lstlisting}[float,basicstyle=\small,caption={Interaction rule that shows SA vulnerability propagation},label=lst:sa-vul-propagation]
vulnerableSA(SA, VulID) :- 
  vulnerableSA(SA_input, VulID),
  wasInputTo(SA_input, T),
  generated(T, SA)

vulnerableSA(SA, VulID) :- 
 vulExists(Host, VulID, SA)
\end{lstlisting}

We introduce the predicate \texttt{vulnerableSA(...)} to encode
vulnerability-inference rules derived from SSC interactions. Consider
\Cref{fig:ssc-1}. Let $sa4$ be a vulnerable Java-language software
artifact (e.g., the Log4J library version 2.14.1, which contains the
Log4Shell vulnerability~\cite{Everson2022}). It \textit{was input} to
the \textit{transformer} $t1$, which generated \textit{software
	artifacts} $sa5$ and $sa6$. Listing \ref{lst:sa-vul-propagation} shows
two rules for vulnerability propagation. The first rule states that a
SA is vulnerable if it was generated by the \textit{transformer} $T$
and $T$ used a vulnerable SA as input. The second states that an SA is
vulnerable if it was declared vulnerable in the initial state (e.g.,
\texttt{vulExists(h2, vulLog4Shell, sa4).}). The rules allow the system
to infer an arbitrarily long chain of SSC vertices that propagate
vulnerable \textit{software artifacts} (i.e.,
\texttt{vulnerableSA(...)} appears on both the left and right sides of
the first rule).

An automated SCA analysis of $sa5$ and $sa6$ would detect the presence
of the library $sa4$ in this case because, in Java, the binary library
dependencies are copied to the resulting binary software package.
However, this is not always the case. For other scenarios where the
dependency is statically linked into the generated binaries, simple
scanning for artifacts will not identify the original libraries. Let
$sa4$ be a vulnerable C-language software artifact (e.g., the OpenSSL
library version 1.0.1, which contains the Heartbleed
vulnerability~\cite{Durumeric2014}) that is compiled, statically
linked, and included by the \textit{transformer} $t1$ in the binary
code of $sa5$ and $sa6$. As in the previous case, the extension would
also infer that $sa5$ and $sa6$ are vulnerable and use this information
to further propagate threats.

\begin{lstlisting}[float,basicstyle=\small,caption={Interaction rule that shows SSC vulnerability propagation as input to MulVal reasoning rules}, label=lst:vul-mulval]
vulExists(Host, VulID, SA, Range, Consequence) :-
  vulnerableSA(SA, VulID),
  wasPresent(SA, Host),
  vulProperty(VulID, Range, Consequence)

networkServiceInfo(Host, SA, Protocol, Port, User) :-
  vulnerableSA(SA, VulID),
  vulNetworkProperty(VulID, Protocol, Port, User),
  wasPresent(SA, Host),
  vulProperty(VulID, remoteExploit, privEscalation)
\end{lstlisting}

The effects of vulnerability propagation in the SSC are perceived when
the affected software artifacts are executed. Then, their
vulnerabilities are ready to be exploited. Listing \ref{lst:vul-mulval}
shows some of the rules that allow MulVal to reason about inferred or
declared vulnerabilities of software artifacts that come from the SSC.
The first inference rule states that if there was a vulnerable software
artifact $SA$ present (i.e., observed to be executing) on host $Host$,
then MulVal's \texttt{vulExists(...)} is true. This allows MulVal rules
to infer its consequences.

In the second inference rule of Listing \ref{lst:vul-mulval}, the
extension signals to MulVal the outcome of the vulnerable software
artifact $SA$ found on host $Host$ by SSC vulnerability propagation.
The effect of the vulnerability is the provision of a network service
(i.e., it is ready to receive network connections) on port $Port$, with
the access privileges of $User$. This allows MulVal to reason about
other conditions (e.g., network access permitted) and generate an
attack path that depends on having the network service available.

\subsection{Propagation of malicious software and asset compromises}

We complement the dynamic and static mechanisms of malicious software
detection~\cite{Polamarasetti2024} with inference. Given a set of known
compromised elements in the initial state, this extension infers its
effects for threat propagation. The resulting attack paths include
inferred malicious software artifacts and asset compromises.

\begin{lstlisting}[float,basicstyle=\small,caption={Interaction rules for SSC compromise propagation},numbers=left,numberstyle=\tiny,label=lst:compromises]
compromisedH(H) :- maliciousSA(SA),  wasPresent(SA, H)
compromisedBE(BE) :- compromisedH(H), hosted(H, BE)
compromisedT(T, BE) :- compromisedBE(BE), executed(BE, T)
maliciousSA(SA) :- compromisedT(T, BE),  generated(T, SA)

compromisedT(T, BE) :- 
  executed(BE, T),
  execBatchCode(BE, SA, User),
  canAccessFile(BE, User, write, SA_build),
  wasBuildToolTo(SA_build, T)

principalCompromised(Victim) :-
  hasAccount(Victim, H, User),
  compromisedH(H)

compromisedH(H) :-  execCode(H, root)	
\end{lstlisting}

Consider \Cref{fig:ssc-1}. Let $sa1$ be a malicious \textit{software
	artifact}. The extension will infer by propagation that $\{h3, be1,
	t1\}$ are compromised and that $\{sa5, sa6\}$ are malicious. Listing
\ref{lst:compromises} shows some of the inference rules for compromise
propagation. On line 1, the rule states that a host $H$ is compromised
if there is a malicious \textit{software artifact} $SA$ executing on
it. On line 2, the rule states that a \textit{build environment} $BE$
is compromised if the \textit{host} $H$ that executed it is
compromised. On line 3, the rule states that a \textit{transformer} $T$
is compromised if it was executed by the compromised \textit{build
	environment} $BE$. On line 4, the rule states that a \textit{software
	artifact} $SA$ is malicious if it was generated by a compromised
\textit{transformer} $T$. In the example of \Cref{fig:ssc-1}, $\{sa5,
	sa6\}$ are malicious because of the rules on lines 1-3.

Line 6 of Listing \ref{lst:compromises} shows an attack path where
malicious code compromises the build tool of a build step (a
\textit{transformer}). The rule states that a \textit{transformer} $T$
is compromised if it was executed by a \textit{build environment} $BE$,
there was a malicious code $SA$ executing on $BE$, with principal
$User$ (see \ref{lst:exec-batch}) and write access to $SA_{build}$,
which was a build tool to $T$. For an example, consider
\Cref{fig:ssc-1}. Let $sa2$ be the only malicious \textit{software
	artifact} in the initial state. Then the extension will infer that
$\{sa5, sa6\}$ (generated by $t1$) are malicious if $sa2$ has
\textit{write} access to $sa3$ (the build tool used by the transformer
$t1$).

On line 12 of Listing \ref{lst:compromises}, we show a rule that
connects SSC threat propagation to MulVal rules. The principal $Victim$
is compromised if it has an account on \textit{host} $H$ which is
compromised according to SSC compromise-propagation rules.

Finally, on line 16 of Listing \ref{lst:compromises}, we show a rule
that connects MulVal inference rules to SSC threat-propagation rules.
It states that the \textit{host} $H$ is also compromised if there is a
successful attack path leading to $H$, according to MulVal rules (i.e.,
\texttt{execCode(...)}). We can now present usage scenarios in
\Cref{sec:usage}.

\section{Detecting real-world SSC attacks}
\label{sec:usage}

\begin{lstlisting}[float,basicstyle=\small,caption={MulVal predicates for defining the initial state},label=lst:example-1]
attackerLocated(internet).
hacl(internet, h1, tcp, 443).
vulExists(h1, 'CVE-2021-41773', httpd).
vulProperty('CVE-2021-41773', remoteExploit, privEscalation).
networkServiceInfo(h1, httpd, tcp , 443 , user_apache).
vulExists(h1, 'CVE-2021-3560', polkit).
vulProperty('CVE-2021-3560', localExploit, privEscalation).
\end{lstlisting}

To use the extension, we encode the SSC graph and the initial state
with logic predicates. The MulVal extension then generates the attack
graph using both the existing MulVal predicates and the new extension
predicates. In the scenarios shown, the attack paths cannot be found
with either MulVal or SSC threat-propagation knowledge alone, because
each predicate set covers a different subset of attack steps. These
differences are illustrated by the color coding in the attack graphs.

\begin{figure}
	\centering
	\includegraphics[width=0.9\textwidth]{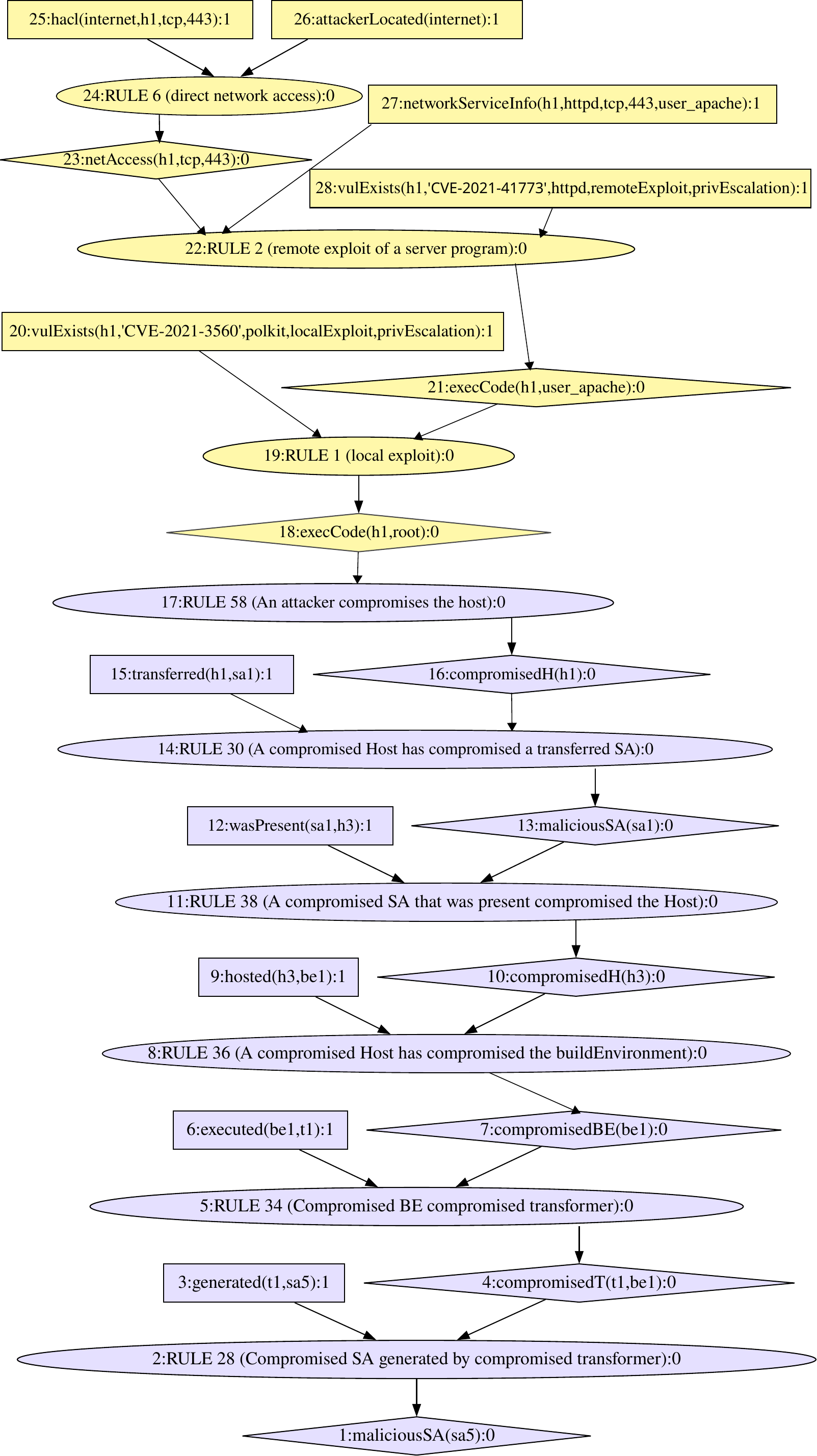}
	\caption{A pruned attack graph generated for the SSC of $sa5$.}
	\label{fig:attack-graph-1}
\end{figure}

In the first scenario, the attack graph is initiated by a cyberattack
on host $h1$ and then SSC threat-propagation rules compute the
subsequent effects. We define the SSC by using predicates from those
introduced in \Cref{sec:ssc-assets} for each edge of \Cref{fig:ssc-1}.
For example: \textit{transferred(h1, sa1)}, \textit{wasPresent(sa1,
	h3)}, \textit{hosted(h3, be1)}, \textit{wasPresent(sa2, be1)},
\textit{executed(be1, t1)}, \textit{wasBuildToolTo(sa3, t1)},
\textit{generated(t1, sa5)}. The initial state is shown in Listing
\ref{lst:example-1}. Apache \textit{httpd} software on host $h1$ is
vulnerable to remote access. Additionally, the \textit{polkit} software
on $h1$ contains a vulnerability that allows privilege escalation. The
resulting attack graph (only shown for $sa5$) appears in
\Cref{fig:attack-graph-1}. Vertices 1-17 (purple) denote new SSC
threat-propagation rules and vertices 18-26 (orange) denote existing
MulVal rules. There is an attack path that leads, in the first steps,
to the root compromise of \textit{host} $h1$. From that point,
\textit{software artifact} $sa1$ is inferred to be malicious, which
leads to the compromise of \textit{host} $h3$, which compromises both
the \textit{build environment} $be1$ and the \textit{software artifact}
$sa4$. Finally, both paths lead to the compromised \textit{transformer}
$t1$, which leads to the malicious \textit{software artifacts} $sa5$
and $sa6$.

\begin{figure}
	\centering
	\includegraphics[width=1.0\textwidth]{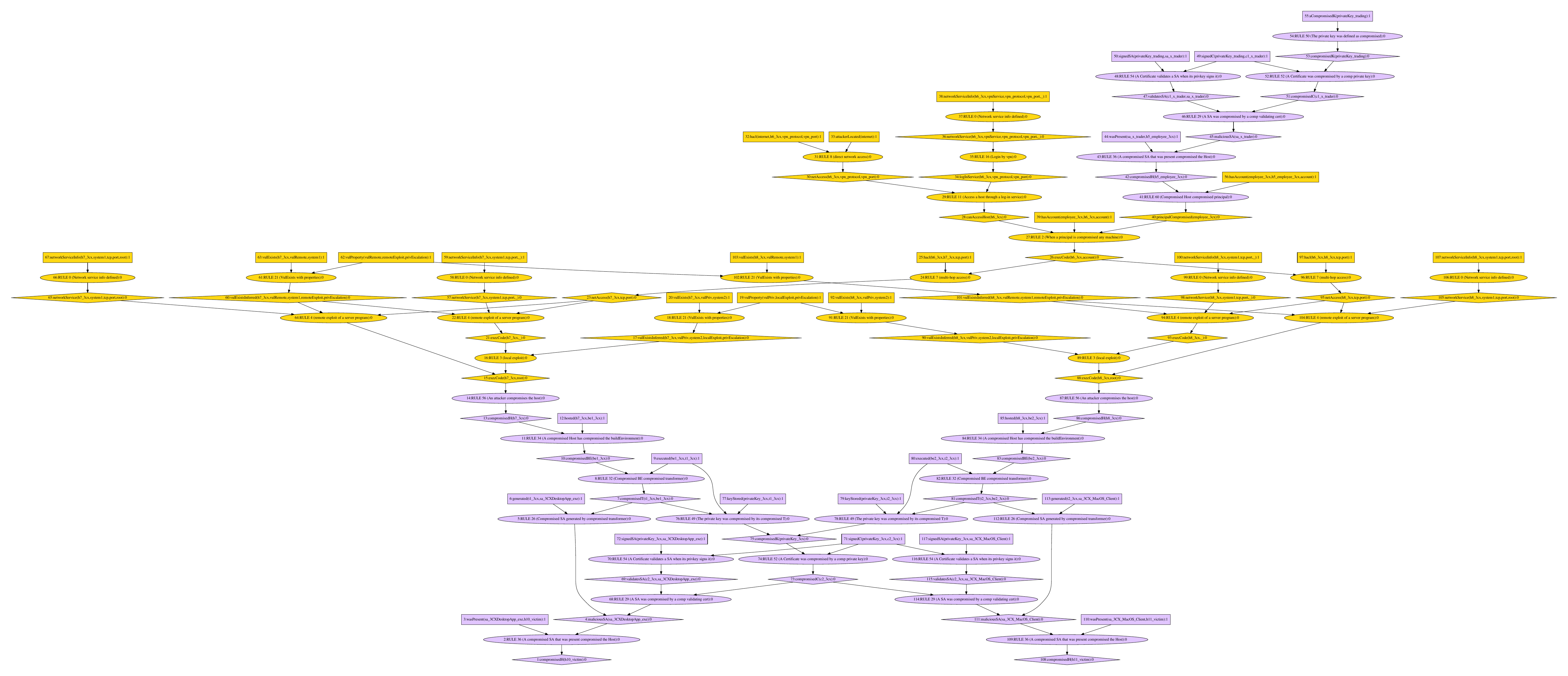}
	\caption{Intertwined rules (orange and purple) for the 3CX attack graph}
	\label{fig:attack-graph-3cx}
\end{figure}

In the second scenario, the attack graph for the 3CX double-SSC attack
is depicted in \Cref{fig:attack-graph-3cx}, color-coded as the previous
scenario. Although individual rule text is unreadable at this scale,
the dense structure and mixed colors convey both the complexity of the
rule set and the tight interdependence between standard attack rules
and the proposed SSC propagation rules. The replication
package~\cite{r_zenodo} contains 20 additional usage scenarios,
including signing-key compromise, build-environment isolation and
escape, and combined SSC attacks.

\section{Discussion}
\label{sec:discussion}

\paragraph{}\textbf{RQ1:} \textit{To what extent is it possible to formalize knowledge of SSC attacks into LAG?} The SSC MulVal extension allows users to account for SSC
attacks, even in long chains of interactions, when generating attack
paths. It can be used to prioritise the resources that appear on the
attack paths for further investigation. In the example shown in
\Cref{sec:usage}, the system infers that the software artifacts
$\{sa5, sa6\}$ at the end of the SSC are malicious. The attack path
begins with an attacker exploiting two vulnerabilities in a remote
host. The example shows that the capability of the logical graph
generator was correctly expanded to also account for the effects of
attacks on the SSC.

\textbf{RQ2:} \textit{To what extent does such a formalism uncover non-trivial attack scenarios?} Because of the inference rules shown, especially in
\Cref{sec:threat-propagation}, threats can be inferred instead of only
detected with external tools. An inferred threat makes it possible to
reason about its effects on the other elements of the SSC and
networked systems, covering complex scenarios. In the best-case
scenario (i.e., the inference is effective), the inferred threats in
the attack paths can be neutralized (e.g., a new firewall rule that
blocks a host from receiving network connections, or a software-artifact
version changed). In the worst-case scenario, the resources on the
attack paths are all false positives. In this case, the effort of
investigating possible compromises is restricted to the resources in
the attack paths, a fraction of all the resources available.

\section{Limitations}
\label{sec:limitations}

We chose a widely-used FOSS tool for LAG generation. However, there
might be other developments that could make the work of integrating SSC
knowledge easier. We chose to implement the extension by only adding
predicates to the base MulVal rules. In this way, they should be
compatible with other extensions. However, if other extensions replace
the original MulVal rules (instead of only adding new rules) it might
cause integration problems. Despite the absence of facts and inference
rules that are specific to FOSS, the need to declare the SSC structure
may be an issue for non-free projects. In this case, the original
MulVal approach can still be used at the expense of accuracy.

\begin{table}
	\caption{Execution time for increasingly larger scenarios.}
	\label{table:execution}
	\centering
	\begin{tabular}[pos]{ |r|r|r|r|}

		\hline
		\textbf{\#Hosts} & \textbf{\#SA} & \textbf{\#Predicates} & \textbf{Time} \\
		\hline
		3K               & 39K           & 4M                    & 53 s          \\
		\hline
		3K               & 183K          & 21M                   & 13 min        \\
		\hline
		6K               & 186K          & 40M                   & 48 min        \\
		\hline
		15K              & 195K          & 100M                  & 4 h 46 min    \\
		\hline
	\end{tabular}
\end{table}

Dealing with very large graphs can pose scalability problems. MulVal
can handle millions of predicates (vertices in the SSC). However, when
full SSC graphs are used—because all software artifacts observed on
each build environment and host must be represented—the reasoning
engine may reach MulVal's limits. We generated scenarios with
increasingly larger SSC graphs to gain insight into the possible
limits. For the experiment we assumed each host and build environment
contains 1,000 to 5,000 unique software packages drawn from a limited
set of operating systems. We executed MulVal with our extension on a
computer equipped with an Intel Core i7-12700H CPU and 32 GB of RAM.
The results are shown in \Cref{table:execution}. The columns list the
number of unique hosts, unique software artifacts, resulting number of
predicates, and total running time for logical attack graph generation.
We stopped after completing the scenario with 100 million predicates
(just under five hours of execution). We observed that the current
implementation uses only a single thread for computation. Expanding
parallelism is one approach to improve the engine's performance.
Another possible solution for supporting even larger SSC graphs is to
partition threat-propagation runs around each software artifact and
cache results in a network-reachable database.

\section{Conclusion}
\label{sec:conclusion}

This paper presents an extension to MulVal by introducing new
predicates and rules to: (i) model SSC assets and their interactions
along which attacks can propagate; (ii) represent assets’ security
status (e.g., vulnerable, malicious, or compromised); (iii) encode
initial knowledge about vulnerable or compromised hosts and software
artifacts; and (iv) model security mechanisms and incorporate them into
SSC threat propagation. This extension captures complex attack
scenarios that combine SSCs and traditional networked-system attacks.
Those scenarios display strong interleaving between attack types,
indicating that threat identification would not be possible with either
reasoning approach alone.

\textit{Future work}. We plan to develop a mechanism for partitioning,
caching, updating, and retrieving partial SSC threat-propagation runs to
guarantee scalability for very large graphs. Another area of work is the
automatic generation of MulVal input rules. We consider the usage of
hardware mechanisms (e.g., Trusted Platform Module) to help
instrumentation logic capture running software-artifact information
during builds.

\subsubsection*{Acknowledgments}
Supported by the industrial chair Cybersecurity for Critical Networked Infrastructures (cyberCNI.fr) with support of the FEDER development fund of the Brittany region, France.

\end{document}